\documentclass[preprint]{aa}
\usepackage{txfonts}
\usepackage{epsfig}
\usepackage{graphicx}

\begin{document}
 \title{Chandra Observation  of   the Big Dipper X 1624--490}
  \subtitle{}

  \author{R. Iaria\inst{1}, G. Lavagetto\inst{1},  
   A. D'A{\'\i}\inst{1}, T. Di Salvo\inst{1},  N. R. Robba\inst{1}}
   \offprints{R. Iaria, iaria@fisica.unipa.it}
   \institute{Dipartimento   di  Scienze   Fisiche   ed  Astronomiche,
              Universit\`a  di  Palermo,  via  Archirafi  36  -  90123
              Palermo,  Italy} 
 \date{Received / Accepted 13/11/2006}

\authorrunning{R. Iaria  et al.} 
\titlerunning  {Chandra Observation  of   the Big Dipper X 1624--490 }

\abstract{ We present the results  of a 73 ks long Chandra observation
  of the dipping source X 1624--490.  During the observation a complex
  dip lasting 4 hours is observed.  We analyse the persistent emission
  detecting,  for  the  first  time  in the  1st-order  spectra  of  X
  1624--490, an absorption line associated to \ion{Ca}{xx}. We confirm
  the  presence  of the  \ion{Fe}{xxv}  K$_\alpha$ and  \ion{Fe}{xxvi}
  K$_\alpha$ absorption lines with a larger accuracy with respect to a
  previous XMM observation.  Assuming that  the line widths are due to
  a bulk motion or a turbulence associated to the coronal activity, we
  estimate  that  the  lines  have  been produced  in  a  photoionized
  absorber  between  the coronal  radius  and  the  outer edge  of  the
  accretion  disk.    } \maketitle

\section{Introduction}
About 10  Low Mass  X-ray Binaries (LMXB)  are known to  show periodic
dips in  their X-ray  light curves  and most of  them also  show X-ray
burst activity.   The dip intensities, lengths and  shapes change from
source  to source,  and, for  the same  source, from  cycle  to cycle.
Among  the  dipping  sources,   X  1624--490  (4U  1624--49)  exhibits
irregular dips  with a long  orbital period of  21 hr (Watson  et al.,
1985) suggesting  that the source  has the largest  stellar separation
and for this  reason it is called the "Big  Dipper".  Dipping is deep,
during the dips the flux in  the band 1--10 keV decreases of 75\% with
respect  to the  persistent  flux.  The  source  also exhibits  strong
flaring in which  the X-ray flux can increase  by 30\% over timescales
of a few  thousand seconds (Church \& Balucinska-Church  1995), but no
X-ray bursts have been observed (e.g., Balucinska-Church et al.  2001,
Smale et al.  2001).  Another peculiarity of X 1624--490 is that it is
the most luminous  dipping source with a 1--30  keV luminosity of $7.3
\times 10^{37}$ erg s$^{-1}$ (Balucinska-Church et al., 2000) assuming
a distance to the source of  15 kpc (Christian \& Swank, 1997).  Jones
\& Watson (1989) studied the spectrum of X 1624--490 taken from a long
{\it  Exosat} observation  of 220  ks  and from  Ginga, modelling  the
continuum emission  with a blackbody plus  a bremsstrahlung component.
Church  \&  Balucinska-Church   (1995)  reanalysed  the  {\it  Exosat}
observation fitting the continuum of X 1624--490 with a blackbody plus
a cutoff power law.  Balucinska-Church et al. (2000) modelled the broad
band BeppoSAX  spectrum of X 1624--490 during  the persistent emission
adopting  a blackbody  plus  a Comptonization  model  (described by  a
cutoff power  law) with a blackbody  temperature of 1.3  keV, a photon
index  of 2,  and a  cutoff  energy of  12 keV  during the  persistent
emission; during the deep  dipping the blackbody component was totally
absorbed  while the  Comptonized component  was little  absorbed.  The
continuum emission was absorbed by neutral matter having an equivalent
hydrogen  column  of $9  \times  10^{22}$  cm$^{-2}$  larger than  the
expected Galactic  one.  This  large value could  be explained  by the
presence of a  dust scattered halo around the  source, demonstrated by
Angelini et al.  (1996).

\begin{figure*}[t]
\resizebox{\hsize}{!}{\includegraphics{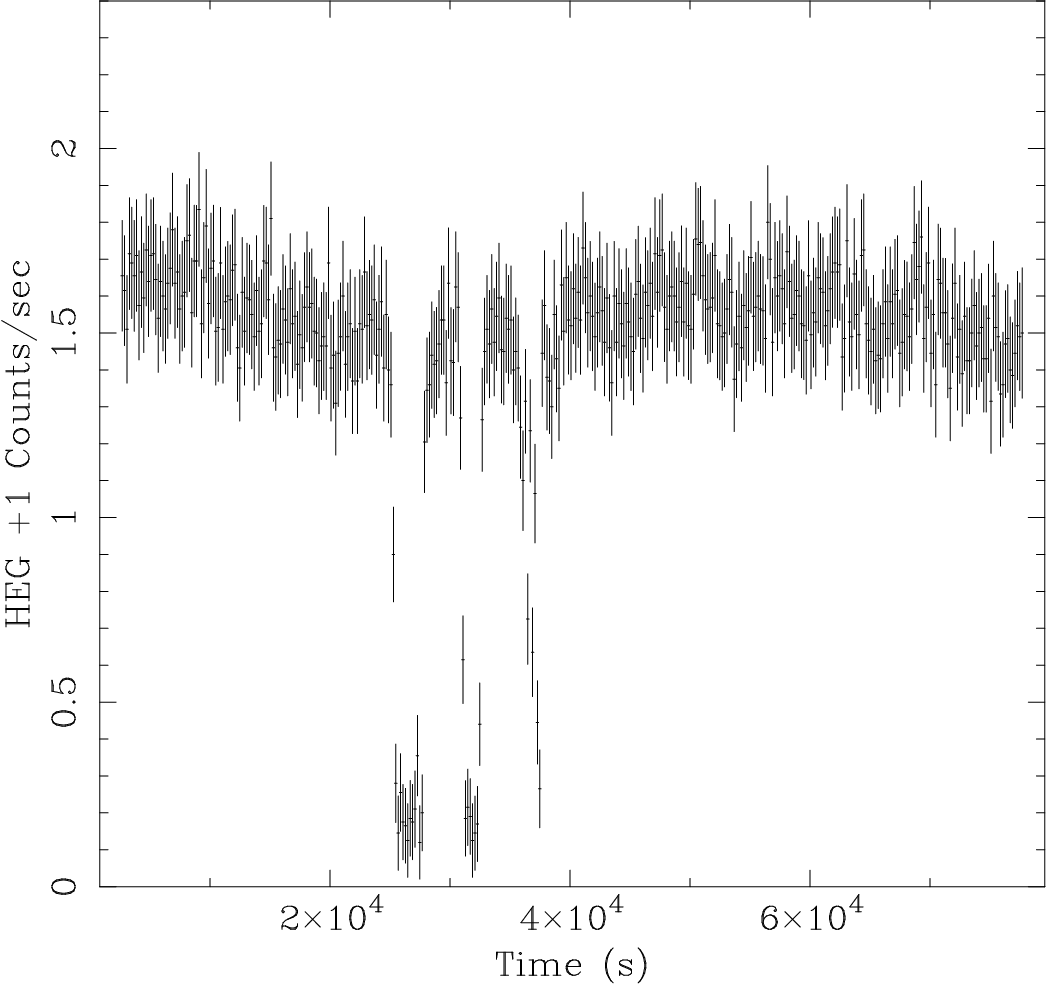} \includegraphics{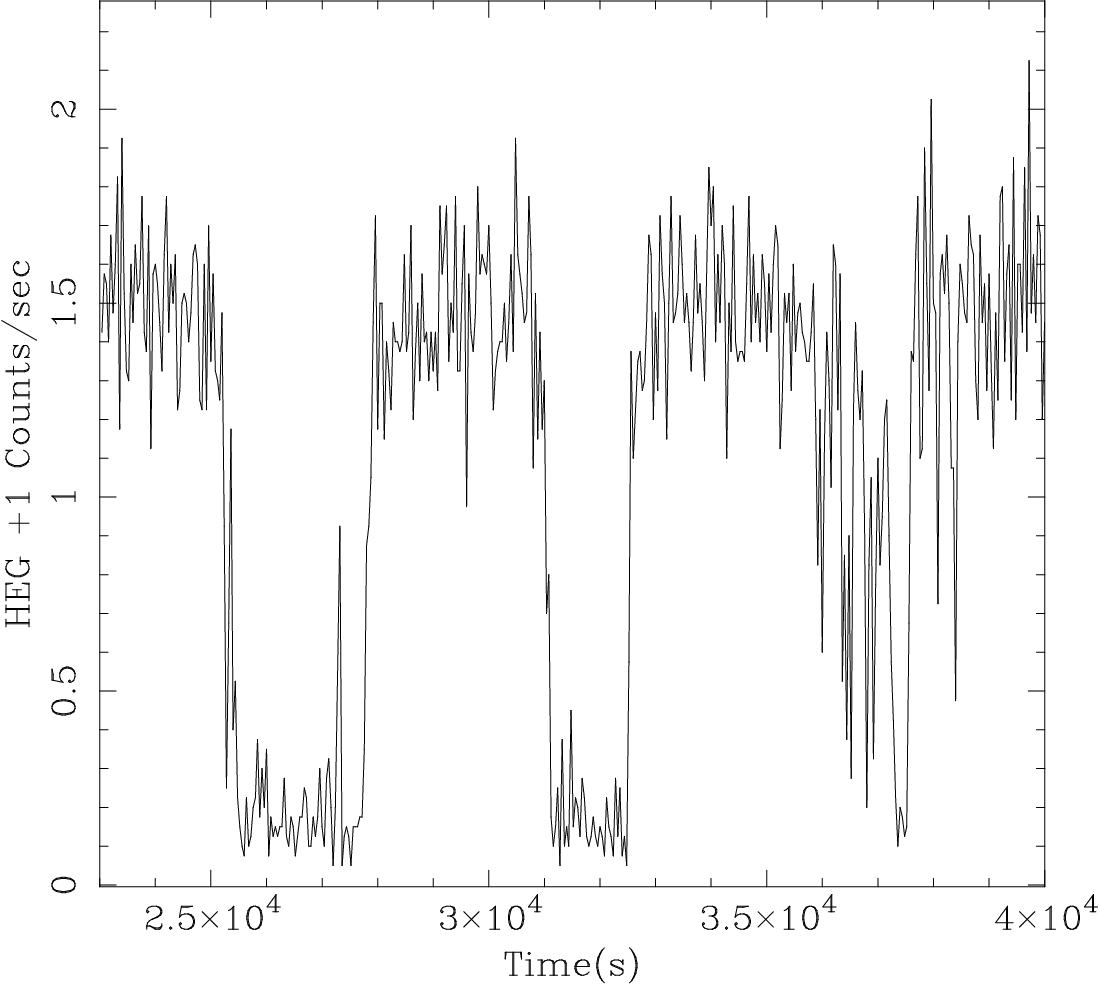}}
\caption{{\bf Left panel:} the 73  ks lightcurve of X 1624--490 with a
  bin  time of 200  s.  The  used events  corresponds to  the positive
  first-order HEG. During the  observation a complex dip lasting $\sim
  4$  hrs was  present.   {\bf  Right panel:}  detail  of the  dipping
  activity, the bin time is 40 s. }
\label{fig1}
\end{figure*}

Parmar et  al.  (2002)  studied X 1624--490  using a 52  ks XMM-Newton
EPIC  observation.  The continuum  of the  persistent emission  in the
2--10  keV EPIC  pn  spectrum was  fitted  using a  blackbody, with  a
temperature of 1.2 keV, plus a power-law component with a photon index
of 2.  The spectrum showed  also narrow features which were identified
as \ion{Fe}{xxv}  K$_\alpha$ and \ion{Fe}{xxvi}  K$_\alpha$ absorption
lines centered at $6.72 \pm 0.03$  and $7.00 \pm 0.02$ keV, with upper
limits on the line widths of 56 and 50 eV, respectively.  Furthermore,
a Fe emission line centered at 6.58 keV with a width and an equivalent
width of 470 and 78 eV, respectively.

Finally we note that  the improved sensitivity and spectral resolution
of Chandra  and XMM-Newton are  allowing to observe  narrow absorption
features, from highly  ionized ions (H-like and He-like),  in a larger
and  larger number of  X-ray binaries.   These features  were detected
in the  micro-quasars GRO J1655--40  (Ueda et al.  1998;  Yamaoka et
al.  2001) and  GRS 1915+105 (Kotani et al.  2000;  Lee et al.  2002).
Recently  the Chandra  High-Energy  Transmission Grating  Spectrometer
(HETGS) observations  of the black hole candidate  H 1743--322 (Miller
et al.  2004) have  revealed the presence of blue-shifted \ion{Fe}{xxv}
and  \ion{Fe}{xxvi} absorption  features  suggesting the  presence of  a
highly-ionized  outflow.    All  LMXBs  those   exhibit  narrow  X-ray
absorption  features are  all known  dipping sources  (see Table  5 of
Boirin  et al.   2004)  except for  GX  13+1. This  source shows  deep
blue-shifted  Fe  absorption features  in  its  HETGS spectrum,  again
indicative of outflowing material (Ueda et al. 2004). We conclude this
brief resume observing that a  recent Chandra spectral analysis of the
dipping source XB 1916--053  showed the presence of several absorption
features  associated to  \ion{Ne}{x}, \ion{Mg}{xii},  \ion{Si}{xiv} and
\ion{S}{xvi}, \ion{Fe}{xxv}  and \ion{Fe}{xxvi}  (see Iaria et  al., 2006a;
Juett \& Chakrabarty, 2006).

In this work we present a spectral analysis of the persistent emission
from the  dipping source X 1624--490  in the 1.7--10  keV energy range
using  a long  73 ks  Chandra observation (Obsid 4559 in the archive of
public Chandra observations).   During the  observation a
complex  dip  lasting  4 hr  was  present,  we  excluded it  from  our
analysis.    We   fitted  the   continuum   emission  adopting   three
statistically  equivalent  models:  a  Comptonized  component  with  an
electron temperature of  1.9 keV, a blackbody plus a  power law, and a
cutoff  power law.   We clearly  detected a  \ion{Fe}{xxvi} K$_\alpha$
absorption line, already observed with the XMM observation; the better
energy resolution of Chandra and the larger statistics allowed to well
determine  its width.   Moreover  we weakly  detected a  \ion{Fe}{xxv}
K$_\alpha$ absorption line  and, for the first time  in the spectra of
this source, a \ion{Ca}{xx}  absorption line.  We discuss the possible
production mechanisms  of these  absorption features and  the physical
geometry of absorbing medium.

\section{Observation} 
X 1624--490 was observed with  the Chandra observatory on 2004 June 04
from  06:26:06 to  2004  June 05  03:46:17  UT using  the High  Energy
Transmission Grating Spectrometer (HETGS).  The observation had a total
integration time  of 73  ks, and was  performed in timed  graded mode.
The HETGS consists  of two types of transmission  gratings, the Medium
Energy Grating  (MEG) and  the High Energy  Grating (HEG).   The HETGS
affords high-resolution spectroscopy from 1.2 to 31 \AA\ (0.4--10 keV)
with a peak spectral  resolution of $\lambda/\Delta \lambda \sim 1000$
at 12 \AA\  for HEG first order.  The  dispersed spectra were recorded
with an array  of six charge-coupled devices (CCDs)  which are part of
the   Advanced   CCD   Imaging   Spectrometer-S   (Garmire   et   al.,
2003)\footnote{See  http://asc.harvard.edu/cdo/about\_chandra for more
  details.}.  We processed the  event list using the standard software
(FTOOLS v6.0.2  and CIAO v3.2 packages)  and computed aspect-corrected
exposure maps for each spectrum.
 
The brightness  of the source required additional  efforts to mitigate
``photon pileup'' effects. A 512  row ``subarray'' (with the first row
= 1) was applied during the observation reducing the CCD frame time to
1.74 s.   Pileup distorts the  count spectrum because  detected events
overlap  and  their  deposited  charges  are  collected  into  single,
apparently more  energetic, events.   Moreover, many events  ($\sim 90
\%$) are  lost as the grades of  the piled up events  overlap those of
highly energetic background particles and  are thus rejected by the on
board software.  We, therefore, ignored the zeroth-order events in our
spectral analysis.  On  the other hand, the grating  spectra were not,
or only  moderately (less  than 10 \%),  affected by pileup.   In this
work  we analysed  the 1st-order  HEG and  MEG spectra.   To precisely
determine the zero-point position in  the image, we estimated the mean
crossing point of the zeroth-order readout trace and the tracks of the
dispersed HEG  and MEG arms.  In  Fig.~\ref{fig1} we showed  the 200 s
bin  time  lightcurve taking  into  account  only  the events  in  the
positive first-order  HEG.  During the  observation was present  a dip
lasting around  4 hrs  and showing a  complex nature of  the occulting
bulge  in  the outer  disk.   The  count  rate during  the  persistent
emission  and the  dipping  episodes  was around  1.5  and 0.15  count
s$^{-1}$, respectively, with  a decrease of the intensity  of 77\%, as
aspected for this source (Church, \& Balucinska-Church, 1995). Finally
we noted  that no  flare activity was  observed during  the persistent
emission.

 \begin{figure}[h]
  \resizebox{\hsize}{!}{\includegraphics{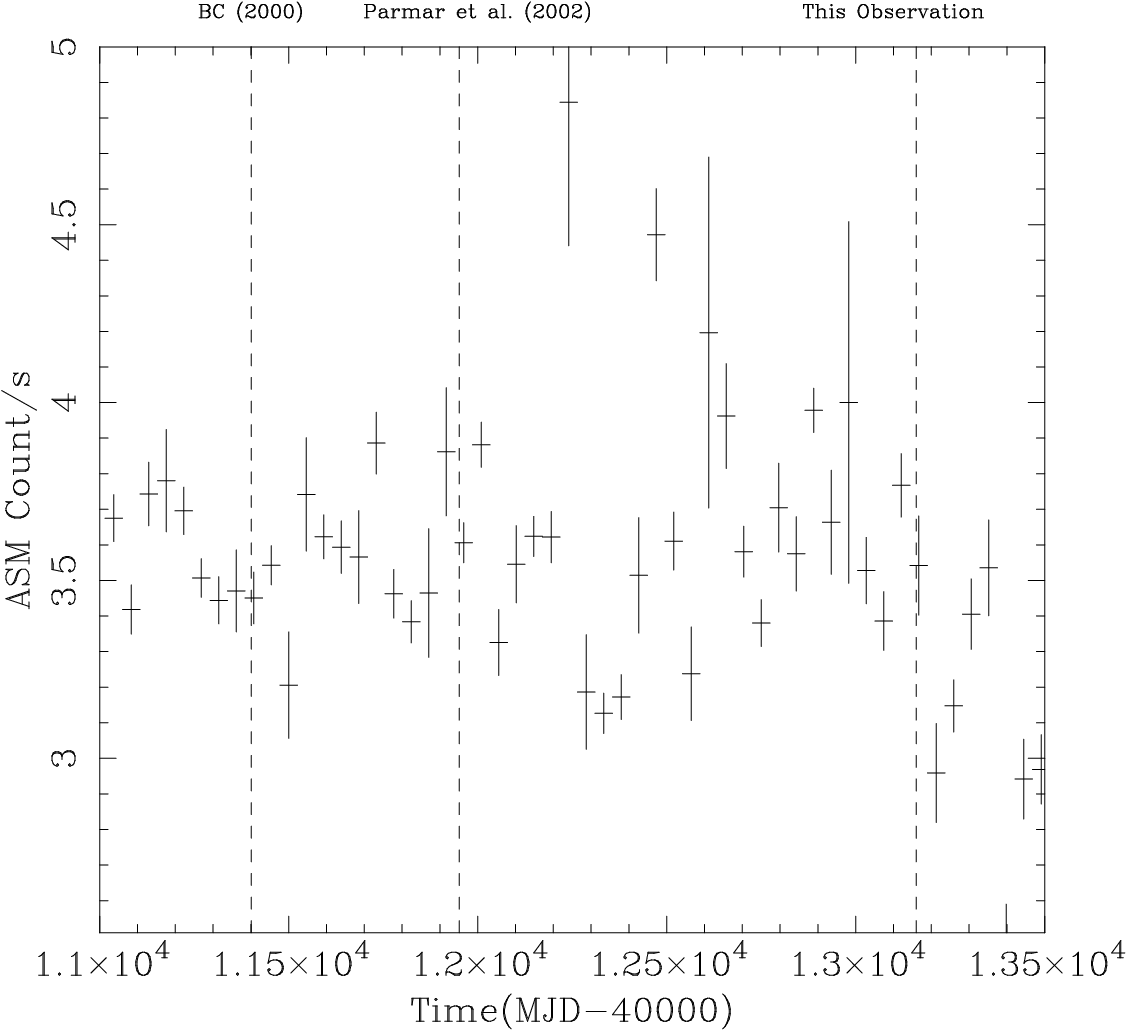}}
  \caption{The RXTE ASM lightcurve  of X 1624--490.  The dashed vertical
    lines  indicate  the  start  times  of  the  BeppoSAX  observation
    (Balucinska-Church et al., 2000),  XMM observation (Parmar et al.,
    2002), and this observation. }
\label{fig2b}
\end{figure}

We analysed  the RXTE  ASM lightcurve of  X 1624--490  observing that
during  the previous BeppoSAX  observation (Balucinska-Church  et al.,
2000), the  previous XMM  observation (Parmar et  al., 2002),  and our
observation  the  ASM  count  rate  was  around  3.5  counts  s$^{-1}$
suggesting a similar intensity of the source.  In Fig.  \ref{fig2b} we
showed  the ASM  lightcurve of  X 1624--490  and indicated  the start
times of the three observations.

\section{Spectral Analysis}
 
Since the lightcurve and the hardness ratio of the source did not show
any significant change during  the persistent emission we selected the
1st-order spectra from the HETGS data obtaining a total effective time
of 65 ks.   Data were extracted from regions  around the grating arms;
to avoid overlapping  between HEG and MEG data, we  used a region size
of  25 and  33 pixels  for the  HEG and  MEG, respectively,  along the
cross-dispersion direction.   The background spectra  were computed by
extracting data above and  below the dispersed flux.  The contribution
from the background is $0.4 \%$  of the total count rate.  We used the
standard CIAO tools to create detector response files (Davis 2001) for
the HEG -1 (MEG -1)  and HEG +1 (MEG +1) order (background-subtracted)
spectra.  After  verifying that the negative and  positive orders were
compatible with each  other in the whole energy  range we coadded them
using  the script  {\it add\_grating\_spectra}  in the  CIAO software,
obtaining the  1st-order MEG spectrum and the  1st-order HEG spectrum.
Finally  we rebinned  the resulting  1st-order MEG  and  1st-order HEG
spectra to 0.015 \AA\ and 0.0075 \AA, respectively.

To  fit the  continuum we  rebinned  the resulting  1st-order MEG  and
1st-order  HEG spectra  to 0.015  \AA\ and  0.0075  \AA, respectively.
After  the extraction  of the  spectra  we analysed  the energy  range
1.7--7 keV  and 1.9--10 keV  for first-order MEG and  first-order HEG,
respectively,  because the  low statistics  below 1.7  keV due  to the
large absorption of neutral matter.  Initially we fitted the continuum
using  a model  composed of  blackbody plus  a power  law  obtaining a
$\chi^2$(d.o.f.)   of  899(1145).  We  found  an absorbing  equivalent
hydrogen  column of  N$_H =  9.9 \times  10^{22}$ cm$^{-2}$,  a photon
index  of  2.2,  a   power-law  normalization  of  0.61,  a  blackbody
temperature of 1.5 keV, and  a blackbody normalization of $1.18 \times
10^{-2}$.   We observed in the residuals the presence of an excess
  between 6 and 7 keV, a  prominent absorption line near 7 keV and two
  absorption lines at  4 and 6.7 keV. We modelled  the excess adding a
  Gaussian  emission  line  and  the  absorption  lines  adding  three
  Gaussian lines with negative  normalizations.  The addition of these
  components improved  the fit  with a $\chi^2$(d.o.f.)   of 812(1133)
  and the  statistical significance of  the lines in unit  of $\sigma$
  was 2.8, 3.2, 2.9 and 5.7 corresponding to the emission line and the
  absorption  lines at 4  keV, 6.7  keV and  7 keV,  respectively.  We
  obtained  an absorbing  equivalent hydrogen  column of  N$_H  = 9.50
  \times  10^{22}$  cm$^{-2}$, a  photon  index  of  2.0, a  power-law
  normalization of  0.41, a blackbody  temperature of 1.39 keV,  and a
  blackbody  normalization  of  $1.11  \times  10^{-2}$.   The  energy
  centroid, the width,  and the equivalent width of  the emission line
  were  6.72  keV, 500  eV,  and  100  eV, respectively.   The  energy
  centroid,  the width,  and  the equivalent  width  of the  prominent
  absorption line were  6.98 keV, 21 eV, and  -22 eV, respectively; we
  identified it as a \ion{Fe}{xxvi} K$_{\alpha}$ absorption line.  The
  two   weak  absorption   lines  were   associated   to  \ion{Ca}{xx}
  K$_{\alpha}$  and  \ion{Fe}{xxv}  K$_{\alpha}$,  respectively.   The
  energy  centroid,  the  width,  and  the  equivalent  width  of  the
  \ion{Ca}{xx} absorption  line were 4.095  keV, $<27$ eV, and  -3 eV,
  respectively.  The  energy centroid,  the width, and  the equivalent
  width of the \ion{Fe}{xxv} absorption  line were 6.71 keV, $<54$ eV,
  and  -9 eV,  respectively.  The  best values  and  the corresponding
  errors  of the  continuum and  of the  lines are  reported  in Tabs.
  \ref{tab_xmm_cont}  and  \ref{tab_xmm_lines}.   Finally, assuming  a
  distance  to  the  source  of  15 kpc  (Christian  \&  Swank,  1997;
  hereafter  we  assume this  value  of  distance  to the  source)  we
  extrapolated an  unabsorbed luminosity of $\sim  7.3 \times 10^{37}$
  ergs s$^{-1}$ in the 0.6--10 keV energy range.
 \begin{figure*}
  \resizebox{\hsize}{!}{\includegraphics{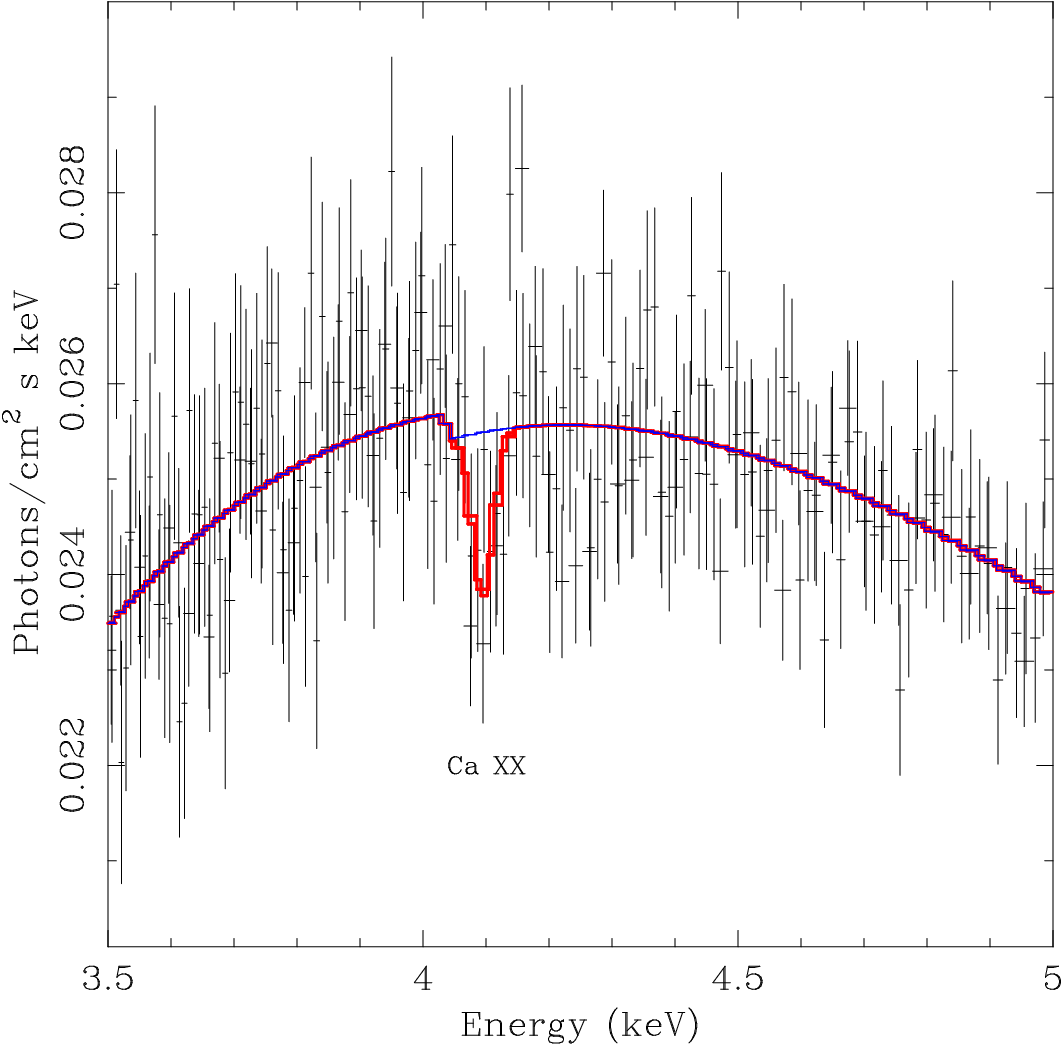}
    \includegraphics{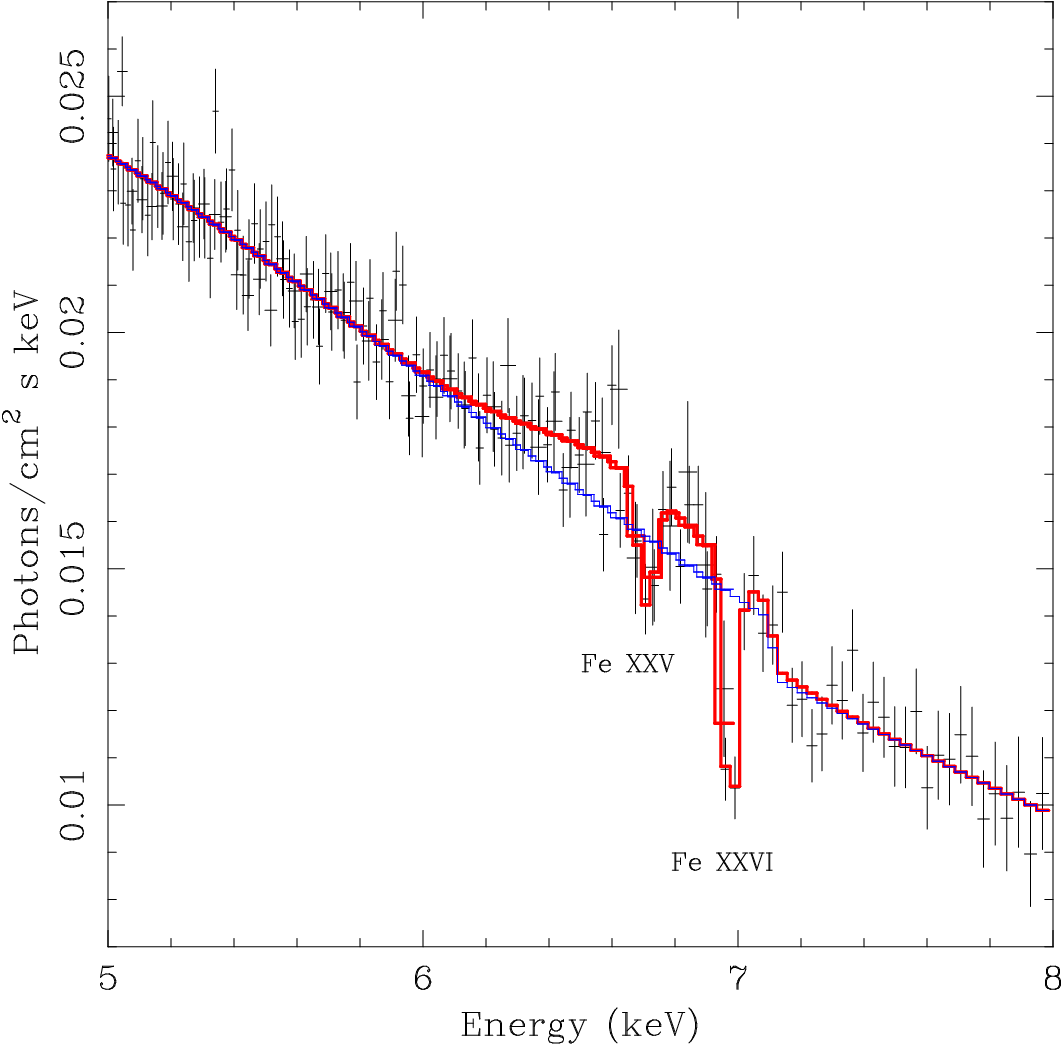}}
  \caption{Chandra/HETGS data of X 1624-490. The Comptonized component
    adopted  as continuum  is indicated  (blue  line) as  well as  the
    continuum modified  by the presence of  the emission and/or absorption
    lines (red  line).  {\bf  Left panel:} the  1st-order MEG  and HEG
    data around 4.1 keV. A  \ion{Ca}{xx} absorption line at $4.095 \pm
    0.015$ keV is present.  {\bf  Right panel:} the 1st-order HEG data
    between 5 and 8 keV.  A Gaussian emission line well fit the excess
    between   6  and  7   keV;  a   \ion{Fe}{xxv}  and   a  prominent
    \ion{Fe}{xxvi}     absorption      line     are     present     at
    $6.709^{+0.019}_{-0.022}$    and   $6.9760   \pm    0.0088$   keV,
    respectively.}
\label{fig3}
\end{figure*} 

   The  spectrum was equivalently  well fitted using as  continuum a
  Comptonized component (Comptt in  XSPEC; Titarchuk, 1994) instead of
  a  blackbody plus  a power  law.  Adopting  the Comptt  component we
  obtained that  the seed-photon temperature k$T_0$ was  0.63 keV, the
  electron temperature k$T_e$  was 1.88 keV, and the  optical depth of
  the Comptonizing  cloud (assuming a spherical geometry)  was 22.  We
  extrapolated  an unabsorbed  luminosity of  $\sim 5  \times 10^{37}$
  ergs  s$^{-1}$  in the  0.6--10  keV  energy  range, this  value  is
  slightly lower than that obtained using a blackbody plus a power law
  because of  the different values  of the absorbing  hydrogen column.
  In this case  the Gaussian emission line is  better constrained; the
  centroid, the  width, the equivalent  width and the  significance of
  the emission  line were  6.64 keV, 280  eV, 49 eV,  and 3.1$\sigma$.
  The parameters  of the absorption  lines are quite similar  to those
  obtained  using the  previous model.   The  results of  the fit  are
  reported in Tabs.  \ref{tab_comptt_cont} and \ref{tab_comptt_lines},
  in  Fig.  \ref{fig3}  we showed  the best-fit  continuum  around the
  \ion{Ca}{xx}   absorption  line   and  the   Fe   K$_\alpha$  region
  overplotting the emission and  absorption lines.  Finally we adopted
  the Comptonized component as  continuum and substituted the Gaussian
  emission line  with an emission  line from a  relativistic accretion
  disk  ({\it diskline}  in XSPEC;  Fabian  et al.,  1989) fixing  the
  inclination angle of  the source at 65$^\circ$, typical  value for a
  dipping source not showing eclipses.   Also in this case we obtained
  statistically  equivalent  parameters   of  the  continuum  and  the
  absorption lines, we obtained an  inner radius $< 10^8$ cm, an outer
  radius $> 1.3 \times 10^8$ cm  and an emissivity of the disk scaling
  as $r^{-2.2}$,  where $r$  is the radial  distance from  the compact
  object. The parameters  of the diskline model were  reported in Tab.
  \ref{tab_dl}.

\begin{table}
\footnotesize
\caption{ \footnotesize \linespread{1} 
  Parameters of the continuum plus  of the Gaussian emission line.  
  The continuum  is composed
  of a blackbody plus  a power-law component. 
  The photoelectric absorption is indicated as
  N$_H$.   Uncertainties are at  90\% confidence  level for  a single
  parameter; upper  limits are  at 95\% confidence  level.  kT$_{BB}$
  and  N$_{BB}$  are,  respectively,  the blackbody  temperature  and
  normalization in  units of L$_{39}  / D^2_{10}$, where  L$_{39}$ is
  the luminosity in units of  10$^{39}$ ergs s$^{-1}$ and D$_{10}$ is
  the  distance in  units of  10  kpc.   N$_{po}$ 
  indicates the normalization
  of  the power  law    in unit  of photons
  keV$^{-1}$  s$^{-1}$ cm$^{-2}$  at  1 keV.  E, $\sigma$, I,  EW, and
  FWHM  indicate  the energy
  centroid,  the width, the  intensity in  units of  photons s$^{-1}$
  cm$^{-2}$, the equivalent width, and the full width half maximum of
  the lines. The  pedex  $line$  indicates the parameters of the 
   the Gaussian emission line.}
\label{tab_xmm_cont}
\centering
 \begin{tabular}{l c   }
\hline
\hline
  Continuum + emission line  &    \\
\hline

$N_{\rm H}$ $\rm (\times 10^{22}\;cm^{-2})$  
& $9.50^{+0.43}_{-0.58}$\\
&       \\
 
kT$_{BB}$ (keV)  
&  $1.39^{+0.11}_{-0.13}$\\      
N$_{BB}$ ($\times 10^{-2}$)
 &   $1.11 \pm 0.13$   \\  
 &\\

photon index 
&  $2.00^{+0.30}_{-0.55}$   \\ 
  
N$_{po}$
&  $0.41^{+0.15}_{-0.20}$   \\
 &\\

E$_{line}$ (keV)
&  $6.72^{+0.26}_{-0.18}$  \\
 
$\sigma_{line}$ (eV)
&  $500^{+130}_{-270}$   \\

I$_{line}$  ($\times 10^{-3}$ cm$^{-2}$ s$^{-1}$)
& $1.7^{+3.1}_{-1.0}$ \\

EW$_{line}$  (eV)
&  $100^{+178}_{-57}$ \\
 
FWHM$_{line}$  (km s$^{-1}$)
&  $\sim 5.3 \times 10^4$ \\
Significance
& 2.8$\sigma$\\
 &\\

L$_{0.6-10 {\rm keV}}$ erg s$^{-1}$  
 &   $7.3 \times 10^{37}$  \\
 &\\
 
$\chi^2$(d.o.f.)
 &  814(1133)  \\
 
\hline
\hline
\end{tabular}
\end{table}

\begin{table}
\footnotesize
\caption{ \footnotesize \linespread{1}  Parameters of the  absorption lines
  using  the continuum reported in Tab. \ref{tab_xmm_cont}. 
  Uncertainties are at  90\% confidence  level for  a single
  parameter; upper  limits are  at 95\% confidence  level.  
  The  pedexes    Ca20, Fe25, and  Fe26 are  referred
  to  the \ion{Ca}{xx}, \ion{Fe}{xxv}, and  
  \ion{Fe}{xxvi} absorption lines, respectively. The parameters are
  defined as in Tab. \ref{tab_xmm_cont}.}
\label{tab_xmm_lines}
\centering
 \begin{tabular}{l c   }
\hline
\hline
Absorption Lines   &    \\
\hline

E$_{{\rm Ca 20}}$  (keV)
& $4.095 \pm 0.015$   \\
 
$\sigma$$_{{\rm Ca 20}}$ (eV)
&  $<27$ \\
 
I$_{{\rm Ca 20}}$ ($\times 10^{-4}$ cm$^{-2}$ s$^{-1}$)
&  $-1.39^{+0.73}_{-0.70}$ \\

EW$_{{\rm Ca 20}}$ (eV)
& $-3.0^{+1.6}_{-1.5}$  \\
 
FWHM$_{{\rm Ca 20}}$ (km s$^{-1}$)
&  $<4700$ \\
Significance
& 3.2$\sigma$\\
&\\

E$_{{\rm Fe 25}}$  (keV)
& $6.711^{+0.019}_{-0.021}$   \\
 
$\sigma$$_{{\rm Fe 25}}$ (eV)
&  $<54$ \\
 
I$_{{\rm Fe 25}}$ ($\times 10^{-4}$ cm$^{-2}$ s$^{-1}$)
&  $-1.7^{+1.0}_{-1.5}$ \\

EW$_{{\rm Fe 25}}$ (eV)
& $-9.1^{+5.3}_{-8.1}$  \\
 
FWHM$_{{\rm Fe 25}}$ (km s$^{-1}$)
&  $<5700$ \\
Significance
& 2.9$\sigma$\\
&\\

E$_{{\rm Fe 26}}$  (keV)
& $6.9764 \pm 0.0090 $  \\
 
$\sigma$$_{{\rm Fe 26}}$ (eV)
&  $21^{+16}_{-19}$ \\
 
I$_{{\rm Fe 26}}$ ($\times 10^{-4}$ cm$^{-2}$ s$^{-1}$)
&  $-3.8^{+1.1}_{-1.4}$ \\

EW$_{{\rm Fe 26}}$ (eV)
& $-21.7^{+6.3}_{-8.1}$  \\
 
FWHM$_{{\rm Fe 26}}$ (km s$^{-1}$)
&  $2100^{+1600}_{-1900}$ \\
Significance
& 5.7$\sigma$\\

\hline
\hline
\end{tabular}
\end{table}
 
\begin{table}
\footnotesize
\caption{ \footnotesize \linespread{1} Parameters of the continuum
  plus  the Gaussian emission line. The continuum is composed
  of a Comptonized Component ({\it Comptt} in XSPEC).  
  The photoelectric absorption is indicated as
  N$_H$.   Uncertainties are at  90\% confidence  level for  a single
  parameter; upper  limits are  at 95\% confidence  level. 
  kT$_0$,  kT$_e$, $\tau$,  and
  N$_{Comptt}$  indicate the  seed-photon  temperature, the  electron
  temperature, the optical depth, and the normalization of the Comptt
  model  of   the  Comptonizing   cloud  around  the   neutron  star,
  respectively. The other parameters are defined as in Tab. 
\ref{tab_xmm_cont}.}
\label{tab_comptt_cont}
\centering
 \begin{tabular}{l c   }
\hline
\hline
Continuum + emission line   &    \\
\hline

$N_{\rm H}$ $\rm (\times 10^{22}\;cm^{-2})$  
& $8.22^{+0.47}_{-0.35}$\\
&       \\

kT$_{0}$ (keV)
  &  $0.61^{+0.14}_{-0.12}$      
  \\

kT$_{e}$ (keV)
  &  $1.882^{+0.181}_{-0.095}$      
  \\

$\tau$ 
  &  $22.0^{+1.2}_{-3.4}$      
  \\

N$_{Comptt}$ 
  &  $0.243^{+0.034}_{-0.031}$     \\
&\\ 

E$_{line}$ (keV)
&  $6.64^{+0.16}_{-0.14}$  \\
 
$\sigma_{line}$ (eV)
&  $280^{+250}_{-120}$   \\

I$_{line}$  ($\times 10^{-3}$ cm$^{-2}$ s$^{-1}$)
& $0.86^{+0.74}_{-0.42}$ \\

EW$_{line}$  (eV)
&  $49^{+44}_{-24}$ \\
 
Significance
& 3.1$\sigma$\\
 &\\
 
L$_{0.6-10 {\rm keV}}$ erg s$^{-1}$  
 &   $5 \times 10^{37}$  \\
 &\\
 
$\chi^2$(d.o.f.)
 &  817(1133)  \\
 
\hline
\hline
\end{tabular}
\end{table}

\begin{table}
\footnotesize
\caption{ \footnotesize \linespread{1}  Parameters of the  absorption lines
  corresponding to  the model reported in Tab. \ref{tab_comptt_cont}. 
  The parameters
  are defined as in  Tabs. \ref{tab_xmm_cont} and \ref{tab_xmm_lines}.}
\label{tab_comptt_lines}
\centering
 \begin{tabular}{l c   }
\hline
\hline
Absorption Lines   &    \\
\hline

E$_{{\rm Ca 20}}$  (keV)
& $4.095 \pm 0.015$     \\
 
$\sigma$$_{{\rm Ca 20}}$ (eV)
& $16.1^{+13.6}_{-7.6}$  \\
 
I$_{{\rm Ca 20}}$ ($\times 10^{-4}$ cm$^{-2}$ s$^{-1}$)
&  $-1.19^{+0.58}_{-0.73}$ \\

EW$_{{\rm Ca 20}}$ (eV)
& $-2.8^{+1.3}_{-1.7}$  \\
 
FWHM$_{{\rm Ca 20}}$ (km s$^{-1}$)
&  $2800^{+1900}_{-1500}$ \\
Significance
& 3.5$\sigma$\\
&\\

E$_{{\rm Fe 25}}$  (keV)
& $6.709^{+0.019}_{-0.022}$   \\
 
$\sigma$$_{{\rm Fe 25}}$ (eV)
&  $<62$ \\
 
I$_{{\rm Fe 25}}$ ($\times 10^{-4}$ cm$^{-2}$ s$^{-1}$)
&  $-1.9^{+1.1}_{-1.5}$ \\

EW$_{{\rm Fe 25}}$ (eV)
& $-10.5^{+5.9}_{-8.1}$  \\
 
FWHM$_{{\rm Fe 25}}$ (km s$^{-1}$)
&  $<6600$ \\
Significance
& 2.7$\sigma$\\
&\\

E$_{{\rm Fe 26}}$  (keV)
& $6.9760 \pm 0.0088 $  \\
 
$\sigma$$_{{\rm Fe 26}}$ (eV)
&  $<35$ \\
 
I$_{{\rm Fe 26}}$ ($\times 10^{-4}$ cm$^{-2}$ s$^{-1}$)
&  $-3.3^{+1.1}_{-1.5}$ \\

EW$_{{\rm Fe 26}}$ (eV)
& $-19.3^{+6.6}_{-8.7}$  \\
 
FWHM$_{{\rm Fe 26}}$ (km s$^{-1}$)
&  $<3600$ \\
Significance
& 4.5$\sigma$\\

\hline
\hline
\end{tabular}
\end{table}
\begin{table}
\footnotesize
\caption{ \footnotesize \linespread{1} Parameters of the 
  {\it diskline} adopting a continuum  composed
  of a Comptonized Component ({\it Comptt} in XSPEC).  
  Uncertainties are at  90\% confidence  level for  a single
  parameter; upper  limits are  at 95\% confidence  level. The
  parameters are defined in the text.}
\label{tab_dl}
\centering
 \begin{tabular}{l c   }
\hline
\hline
Continuum + emission line   &    \\
\hline

E$_{diskline}$ (keV)
&  $6.62^{+0.15}_{-0.18}$  \\
 
Emissivity slope
&  $-2.20^{+1.05}_{-0.43}$   \\

R$_{in}$ (cm)
& $< 10^8$ \\

R$_{out}$ (cm)
& $> 1.3 \times 10^8$ \\

I$_{diskline}$  ($\times 10^{-3}$ cm$^{-2}$ s$^{-1}$)
& $1.27^{+1.01}_{-0.87}$ \\

EW$_{line}$  (eV)
&  $74^{+60}_{-51}$ \\

\hline
\hline
\end{tabular}
\end{table}
 
\section{Discussion}
We  have  analyzed a  65  ks  Chandra  observation of  the  persistent
emission  from  X~1624--490  performing  a spectral  analysis  of  the
persistent  emission using  the 1st-order  MEG and  HEG  spectra.  The
continuum  emission  was equivalently  well  fitted  adopting a  model
composed  of either  a blackbody  plus a  power law  or  a Comptonized
component (Comptt;  Titarchuk, 1994).  A  power law was  chosen rather
that  a  cutoff   power  law  since  the  cutoff   energy  of  12  keV
(Balucinska-Church et al.,  2000) was too high to  be estimated in the
Chandra  energy range; the  same model  was adopted  by Parmar  et al.
(2002) to  fit the persistent  emission continuum of the  source using
XMM  EPIC data.   We  obtained similar  best-fit  parameters to  those
obtained by Balucinska-Church et al.  (2000) and Parmar et al.  (2002)
obtaining an extrapolated unabsorbed  luminosity of $7.3 \times 10^{37}$
erg s$^{-1}$ in  the 0.6--10 keV energy range.

Using as  continuum a  blackbody plus  a power law  we noted  that the
best-fit value of the absorbing equivalent hydrogen column N$_H$ was $
9.5  \times 10^{22}$ cm$^{-2}$,  compatible with  the value  obtained by
Balucinska-Church et al.  (2000), on the other hand using as continuum
the Comptonized  component N$_H$ was $ 8.2  \times 10^{22}$ cm$^{-2}$.
Although the column density of X 1624--490 is high, radio measurements
indicated a  lower value  of $2 \times  10^{22}$ cm$^{-2}$  (Dickey \&
Lockman 1990; Stark et al.  1992),  so that could be possible that the
high value of N$_H$ can be  due to absorption intrinsic to the source,
as already suggested  by Angelini et al. (1997)  and Diaz-Trigo et al.
(2005).
 
We noted that the model composed of a blackbody component plus a power
law  in the  energy range  0.6--10 keV  could be  compatible  with the
Birmingham  Model  (Church  \&  Balucinska-Church,  2004),  where  the
blackbody  emission is  produced by  a thermalized  emission  from the
neutron star and the cutoff power law is the Comptonized emission from
an extended accretion disk corona (ADC) above the accretion disk.
  However   the  blackbody  radius  obtained   from  the  best-fit
  parameters was $R_{BB}  = 7.2 \pm 1.8$ km,  smaller than the typical
  neutron star  radius of 10 km,  suggesting that part  of the thermal
  emission is occulted.
The model composed  of a Comptonized component indicated  that we were
not observing  the primary emission  from the neutron star  and/or the
inner accretion disk.  Under  this hypothesis the Comptonized emission
could be produced in an extended accretion disk corona (ADC) embedding
the accretion disk and the neutron star, in the following we discussed
this scenario.   We assumed a spherical geometry  of the Comptonizing
cloud although we cannot exclude a different geometry.  Following in't
Zand  et  al.   (1999)  we   calculated  the  radius,  R$_W$,  of  the
seed-photon  emitting region  using  the parameters  reported in  Tab.
\ref{tab_comptt_cont}; we obtained  R$_W = 20 \pm 9$  km implying that
the seed  photons for the  Comptonization might come from  the neutron
star surface and/or the inner region of the accretion disk.  According
to White  \& Holt (1982) the  ADC radius R$_{ADC}$ can  be obtained by
R$_{ADC} = (M_{NS}/M_{\odot})T_7^1$ R$_{\odot}$, where $M_{NS}$ is the
mass of  the neutron  star, $M_{\odot}$ and  R$_{\odot}$ are  mass and
radius of the Sun, and $T_7$ is the ADC temperature in units of $10^7$
K. Using  as ADC temperature the electron  temperature k$T_e$ obtained
by  the Comptonized  component and  $M_{NS}= 1.4  M_{\odot}$  we found
R$_{ADC} \simeq  7 \times 10^{10}$  cm.  More precise measures  of the
R$_{ADC}$  made   measuring  the  dip   ingress  time  by   Church  \&
Balucinska-Church (2004) indicated the the R$_{ADC}$ of X 1624--490 is
between 5.8 and 7.2 $\times 10^{10}$ cm, compatible with our result.

 We observed  an emission line between 6 and 7  keV.  We fitted it
  using    a    Gaussian    profile    inferring   a    centroid    of
  $6.64^{+0.16}_{-0.14}$ keV, a width of $280^{+250}_{-120}$ eV and an
  equivalent    width    of     $49^{+44}_{-24}$    eV    (see    Tab.
  \ref{tab_comptt_cont}).  This  feature was already  observed with an
  ASCA observation (see  Tab. 5 in Asai et al.; 2000)  and with an XMM
  observation by Parmar  et al.  (2002).  Parmar et  al.  (2002) found
  that the energy, the width  and the equivalent width of the emission
  line   were  $6.58^{+0.07}_{-0.04}$   keV,  $470\pm   70$   eV,  and
  $78^{+19}_{-6}$ eV,  respectively.  Note that the  parameters of the
  Gaussian emission line were  compatible with those obtained from the
  XMM observation.
Broad Fe  lines have been detected  using XMM from  other neutron star
systems  that  show highly-ionized  absorption  features  such as  MXB
1658-298 (FWHM$=1.4^{+0.3}_{-0.4}$ keV, Sidoli  et al.  2001), GX 13+1
(FWHM$=  1.9 \pm  0.5$  keV, Sidoli  et  al.  2002),  and 4U  1323--62
(FWHM$=2.0^{+0.6}_{-0.4}$ keV,  Boirin et al. 2005),  however this was
the first source observed by Chandra  to show both a broad Fe line and
narrow  absorption   features.   We  immediately   excluded  that  the
broadening of  the line could have  a thermal origin  since the needed
temperature  of the plasma  should be  larger than  10 MeV.   The most
probable origin of the observed  emission line is either the accretion
disk  (in this  case, the  large width  of the  line would  be  due to
Doppler and relativistic  smearing effects) or the ADC  (in this case,
the large width  of the line would be due  to Compton broadening).  We
investigated the possibility that  the line was produced by reflection
in an accretion disk substituting the Gaussian component with the {\it
  diskline}  model in XSPEC  (see Tab.   \ref{tab_dl}).  Unfortunately
the low  statistics and  the overwhelming continuum  did not  allow to
well  determine the  model parameters,  for this  reason we  fixed the
source inclination angle to  65$^{\circ}$, typical value for a dipping
source not  showing eclipses.    We found  an emissivity  slope of
  $-2.20^{+1.05}_{-0.43}$, and  estimated that the  required inner and
  outer radius of the disk were $<480$ and $>640$ gravitational radius
  R$_g$, for a neutron star mass of 1.4 $M_{\odot}$ corresponding to a
  inner and outer radius of the disk $< 10^8$ and $>1.3 \times 10^{8}$
  cm,  respectively.   These values  gave  few  information about  the
  region of the disk in which the line was produced although the upper
  limit on the inner radius indicated that the line should be produced
  in the region  of the disk covered by  the ADC.  Alternatively, the
Compton scattering in  the ADC could explain the  large width $\sigma$
of  the  broad Gaussian  emission  line.   Detailed calculations  give
$\sigma  = 0.019  E \tau_T  (1 +  0.78 kT_e)$,  where $\tau_T$  is the
Thompson optical depth and kT$_e$ is in units of keV (Kallman \& White
1989; see  also Brandt  \& Matt 1994).   Assuming an  average electron
temperature of kT$_e = 1.88$ keV,  as derived from the fit of our data
to the  Comptonization model, we could  explain the width  of the iron
line assuming a Thomson optical depth of $\tau_T = 0.2^{+0.2}_{-0.1}$.
Therefore,  it is possible  that the  line was  produced in  the outer
region  of the  ADC, where  the optical  depth might  be lower  (if we
assume  that  the ADC  temperature  profile  remains constant).   This
result  was  not  unreasonable  since  any contribution  to  the  line
produced inside  the Comptonizing region, where the  optical depth can
be  as  high  as   20  (see  Tab.   \ref{tab_comptt_cont}),  would  be
completely smeared by Comptonization.  Therefore, as expected, we only
observed  that  part  of the  line  that  was  produced in  the  outer
Comptonizing region.  Unfortunately, we could not discriminate between
the two possible origins of  the iron line and its broadening proposed
above,  it  seems  more  appropriate  to consider  that  both  Compton
scattering and  the Doppler shift due  to the Keplerian  motion of the
accretion disk worked  simultaneously in the ADC to  produce the broad
Fe line.

   We  detected   two  narrow   absorption  lines   associated  to
  \ion{Ca}{xx}   K$_\alpha$  and   \ion{Fe}{xxv}  K$_\alpha$,   and  a
  prominent  narrow  absorption   line  associated  to  \ion{Fe}{xxvi}
  K$_\alpha$.  The \ion{Fe}{xxvi} line energy did not fit with the lab
  energies.  We believe that this was not due to a physical effect but
  to   a   systematic  error   associated   to   the  uncertainty   of
  0.6\arcsec\footnote{See     http://cxc.harvard.edu/cal/ASPECT/celmon/
    for more  details.}  of the source position  which also identifies
  the  zero of  the grating  arms.  Considering  the  systematic error
  associated  to the  absolute wavelength  accuracy (i.e.   $\pm 0.006
  {\rm \AA}$ and $\pm 0.011  {\rm \AA}$ for HEG and MEG, respectively)
  the  \ion{Fe}{xxvi} line energy  was fully  consistent with  the lab
  energy.  The \ion{Fe}{xxv}  and \ion{Fe}{xxvi} absorption lines were
  already  observed  by  Parmar  et  al.  (2002),  in  that  case  the
  corresponding  upper limits of  the line  width were  50 and  56 eV,
  respectively,  and the  equivalent widths  were $-7.5^{+1.7}_{-6.3}$
  and  $-16.6^{+1.9}_{-5.9}$  eV,  respectively.   We found  that  the
  \ion{Fe}{xxv} and \ion{Fe}{xxvi} line  width were $<62$ and $<35$ eV
  and the  corresponding equivalent widths  were $-10.5^{+5.9}_{-8.1}$
  and $-19.3^{+6.6}_{-8.7}$ eV (see Tab.  \ref{tab_comptt_lines}).  We
  note  that the  higher energy  resolution of  Chandra allowed  us to
  better  constrained the  energy  centroid of  the \ion{Fe}{xxv}  and
  \ion{Fe}{xxvi} absorption  lines that were $6.709^{+0.019}_{-0.022}$
  and $6.9760  \pm 0.0088$ keV,  respectively.  Parmar et  al.  (2002)
  obtained  the \ion{Fe}{xxv}  and \ion{Fe}{xxvi}  energy  centroid at
  $6.72 \pm 0.03$ and $7.00  \pm 0.02$ keV, respectively. We concluded
  that the \ion{Fe}{xxv}  and \ion{Fe}{xxvi} energy centroids reported
  in this work were fully consistent with the corresponding rest-frame
  values,  It was  not  true for  the  \ion{Fe}{xxvi} energy  centroid
  obtained  from  the  XMM  observation  (see Parmar  et  al.,  2002),
  indicating  a   possible  blue-shift  of  this   line  although  the
  \ion{Fe}{xxv} absorption line did not show blue-shift.

Parmar  et  al.   (2002) detected  two  weak  absorption
features at 7.83 and 8.28 keV associated to \ion{Ni}{xxvii} K$_\alpha$
and \ion{Fe}{xxvi}  K$_\beta$, respectively, nevertheless  the authors
also suggested  that these features could have  an instrumental origin
since  they  were  observed  in  a  spectral  region  where  the  EPIC
calibration  was  still relatively  uncertain  at  the  time of  their
analysis.   We  did not  detect  absorption  features  which could  be
associated   to  \ion{Fe}{xxvi}   K$_\beta$   or  to   \ion{Ni}{xxvii}
K$_\alpha$  but we  could not  exclude  that this  is due  to the  low
effective area of the 1st-order HEG above 7 keV.

 We estimated the \ion{Ca}{xx}, \ion{Fe}{xxv} and \ion{Fe}{xxvi} column
densities  using the relation:  $$ \frac{W_\lambda}{\lambda}=\frac{\pi
  e^2}{m_e c^2}  N_j \lambda  f_{ij}=8.85 \times 10^{-13}  N_j \lambda
f_{ij} $$ where $N_j$ is  the column density for the relevant species,
$f_{ij}$  is the  oscillator strength,  $W_\lambda$ is  the equivalent
width  of the  line, and  $\lambda$ is  the wavelength  in centimeters
(Spitzer 1978, p.  52).  Adopting  $f_{ij} = 0.798$, $f_{ij} = 0.416$,
and   $f_{ij}  =   0.416$  for   \ion{Fe}{xxv},   \ion{Fe}{xxvi},  and
\ion{Ca}{xx}  respectively (see  Verner et  al., 1996),  and  the best
parameters reported in  Tab.  \ref{tab_comptt_lines}, we found N$_{\rm
  Fe  XXV} \simeq  1.2 \times  10^{17} $  cm$^{-2}$, N$_{\rm  Fe XXVI}
\simeq 4  \times 10^{17}  $ cm$^{-2}$, and  N$_{\rm Ca XX}  \simeq 6.1
\times  10^{16} $  cm$^{-2}$.   We noted  that  the \ion{Fe}{xxv}  and
\ion{Fe}{xxvi} column densities were similar to those estimated by Lee
et  al.  (2002) for  GRS 1915+105.   Assuming an  ionization continuum
consisting  of a power  law with  $\Gamma =  2$ (Kallman  \& Bautista,
2001) and  using the  ratio between N$_{\rm  Fe XXVI}$ and  N$_{\rm Fe
  XXV} $ we inferred that log$(\xi) \simeq 4.1$ erg cm s$^{-1}$.  This
value  was  similar than  those  recently  obtained  from the  Chandra
analysis of  the dipping source XB 1916--053  (log$(\xi) \simeq 4.15$;
Iaria  et al.,  2006a; Juett  \& Chakrabarty,  2006) and  XB 1254--690
(log$(\xi)   \ga    4.15$;   Iaria   et    al.,   2006b).  
We also noted that the temperature associated to a photoionized plasma
having a  ionization parameter log($\xi$)  of 4.1 erg cm  s$^{-1}$ was
$2.4  \times 10^6$  K.  Finally, we  note  the inferred  value of  the
ionization parameter  also explained the  \ion{Ca}{xx} absorption line
observed in the spectrum and never observed before.

We investigated the nature of the absorption line widths.  A plausible
scenario,  already  adopted  to   describe  the  line  widths  of  the
absorption  lines detected  in  the Chandra  spectrum  of the  dipping
sources XB 1916--053 (Iaria et  al., 2006a) and XB 1254--690 (Iaria et
al., 2006b), is  that the lines were broadened by  some bulk motion or
supersonic  turbulence  with a  velocity  below  3500  km s$^{-1}$  as
indicated    by    the    FWHMs     of    the    lines    (see    Tab.
\ref{tab_comptt_lines}).  Assuming  that the mechanism  generating the
turbulence  or bulk motion  was due  to the  presence of  the extended
corona we can infer some information about where the absorption lines
were produced.  Coronal models tend to have turbulent velocities which
are locally  proportional to the virial or  rotational velocity (Woods
et al.,  1996).  At $ \sim 7  \times 10^{10}$ cm, the  ADC radius, the
virial velocity should be 520 km/s, considering a neutron star of $1.4
M_{\odot}$.   This velocity  was compatible  with the  values obtained
from the FWHMs of the detected absorption lines and suggested that the
absorption  lines could be  produced in  a photoionized  plasma highly
ionized between the  the external region of the ADC  and the disk edge
($\sim 1 \times 10^{11}$ cm, see Church \& Balucinska-Church, 2004).

\section{Conclusion} 

We  studied the  persistent emission  of X  1624--490 using  a  65 ks
Chandra exposure time using the 1st order MEG and HEG spectra in the
1.7--10 keV energy range.

We  fitted the continuum  adopting a  Comptonizing component  which we
interpreted as  emission from an  extended corona above  the accretion
disk,  although   a  blackbody   plus  power  law   was  statistically
equivalent.  The unabsorbed luminosity in the 0.6--10 keV energy range
was $ 5  \times 10^{37}$ erg s$^{-1}$.  We observed  the presence of a
broad Fe emission line  at 6.64 keV with a FWHM of  0.7 keV.  We could
not discriminate the mechanism  producing the broadening of this line.
Under the hypothesis that it  was produced by Doppler and relativistic
smearing effects in  the accretion disk we deduced  that the region of
the disk where  the line had origin was covered  by the accretion disk
corona,  while  under  the  hypothesis  that the  line  broadening  is
producing by Compton  scattering in the ADC we  inferred that the line
should be produced in the  outer region of the corona. Nevertheless we
could not exclude that the broadening  of the emission line was due to
a combination of both these effects.

In the spectrum  we observed three narrow absorption  features that we
associated  to \ion{Ca}{xx},  \ion{Fe}{xxv},  and \ion{Fe}{xxvi}.   We
inferred that  these  absorption lines
were  produced  in a  photoionized  plasma  with ionization  parameter
log($\xi$) of 4.1  erg cm s$^{-1}$, resulting in  a plasma temperature
of $2.4 \times 10^6$ K.   We estimated that the absorption line widths
could  be  compatible with  a  broadening  caused  by bulk  motion  or
turbulence  connected to  the coronal  activity, produced  between the
external ADC and the disk edge ($10^{11}$ cm).

 \begin{acknowledgements}

  This work  was partially  supported by  the Italian
Space  Agency   (ASI)  and  the  Ministero   della  Istruzione,  della
Universit\'a e della Ricerca (MIUR).
\end{acknowledgements}


\begin{thebibliography}{}

\bibitem{}
Angelini, L., Parmar, A. N., \& White, N. E. 1996, Proc. IAU
   Colloquim 163, ASP Conf. series, 121, 685
\bibitem{}
Asai, K., Dotani, T., Nagase, F., \& Mitsuda, K., 2000, ApJS,
   131, 571
\bibitem{}
Balucinska-Church, M., Humphrey, P. J., Church, M. J., \& Parmar,
A. N.,  2000, A\&A, 360, 583
\bibitem{}
Balucinska-Church, M., Barnard, R., Church, M. J., \& Smale, A. P.,
2001,  A\&A, 378, 847
\bibitem{}
Boirin, L., Parmar, A. N., Barret, et al.,  2004, A\&A, 418, 1061
\bibitem{}
Boirin, L., Mendez, M., Diaz Trigo, M., et al.,  2005, A\&A, 436, 195
  \bibitem{}
Brandt, W. M., \& Matt, G. 1994, MNRAS, 268, 1051
  \bibitem{}
Christian, D. J., \& Swank, J. H. 1997, ApJS, 109, 177
  \bibitem{}
 Church, M. J.  \& Balucinska-Church, M.,   1995, A\&A 300, 441
  \bibitem{}
 Church, M. J.  \& Balucinska-Church, M., 2004, MNRAS, 348,  955
  \bibitem{}
Davis, J. E. 2001, ApJ, 562, 575 
\bibitem{}
 Diaz Trigo, M.,  Parmar, A. N.,  Boirin , L., et al., 2005, A\&A,
 445, 179
 \bibitem{}
Dickey, J. M.,  \& Loekman, F., J., 1990, ARA\&A, 28, 215
 \bibitem{}
Fabian et al., 1989, MNRAS, 238, 729 
 \bibitem{}
Garmire, G. P., Bautz, M. W., Ford, et al., 2003, Proc. SPIE, 4851, 28 
 \bibitem{}
Iaria, R., Di Salvo, T., Lavagetto G.,  Robba, N. R.,  Burderi, L.,
2006a,  ApJ, 647, 1341  
 \bibitem{}
Iaria, R., Di Salvo, T., Lavagetto G.,  Robba, N. R.,  Burderi, L.,
2006b,  submitted to A\&A
\bibitem{}
in't Zand, J. J. M., et al. 1999, A\&A, 345, 100
\bibitem{}
Jones, M.H., Watson, M.G., 1989, Proc. of 23rd ESLAB Symposium, Bologna
\bibitem{}
Juett, A. M., \& Chakrabarty, D., 2006, ApJ, 646, 493  
\bibitem{}
Kallman, T., \& White, N. E. 1989, ApJ, 341, 955
\bibitem{}
Kallman, T., \& Bautista, M., 2001, ApJS, 133, 221
\bibitem{}
Kotani, T., Ebisawa, K., Dotani, T., et al., 2000, ApJ, 539, 413
 \bibitem{}
Lee, J. C., Reynolds, C. S., Remillard, R., et al., 2002, ApJ, 567, 1102
 \bibitem{}
Miller, J. M., Raymond, J., Homan, J., et al., 2004, ArXiv
 Astrophysics e-prints, astro-ph/0406272
  \bibitem{}
Parmar, A. N., Oosterbroek, T., Boirin, L., \& Lumb,  D., 2002, A\&A,
386, 910
  \bibitem{}
Sidoli, L., Oosterbroek, T., Parmar, A. N., Lumb, D., \&
  Erd, C., 2001, A\&A, 379, 540
  \bibitem{}
Sidoli, L., Parmar, A. N., Oosterbroek, T., \& Lumb, D.,
  2002, A\&A, 385, 940
   \bibitem{}
Smale, A. P., Church, M. J., \& Balucinska-Church, M.,  2001,
  ApJ, 550, 962 
\bibitem{}
Stark, A.A., Gammie, C.F., Wilson, R.W., et al., 1992, ApJS 79, 77
\bibitem{}
Titarchuk, L., 1994, ApJ, 434, 570
\bibitem{}
Ueda, Y., Inoue, H., Tanaka, Y., et al., 1998, ApJ, 492, 782
\bibitem{}
Ueda, Y., Murakami, H., Yamaoka, K., et al.,  2004, ApJ, 609, 325
  \bibitem{}
 Verner, D. A., Verner, E. M., Ferland, G. J., 1996, At. Data Nucl. 
Data Tables,   64, 1
  \bibitem{}
Watson, M.G., Willingale, R., King, A.R., Grindlay, J.E.,
  Halpern, J., 1985, IAU Circ. 4051
  \bibitem{}
White, N. E., \& Holt, S. S., 1982, ApJ, 257, 318
 \bibitem{}
Woods, D. T.,  Klein, R. I., Castor, J. I., et al., 1996, ApJ, 461, 767 
\bibitem{}
Yamaoka, K., Ueda, Y., Inoue, H., et al., 2001, PASJ, 53, 179
 


\end{thebibliography}
\end{document}